# Convective Flow of Sisko Fluid over a Bidirectional Stretching Surface

*Asif Munir[1], Azeem Shahzad and Masood Khan*

*Department of Mathematics, Quaid-i-Azam University, Islamabad* 44000*,
Pakistan*

**Abstract:**
*The present investigation focuses the flow and heat transfer characteristics of the steady three-dimensional Sisko fluid flow driven by a bidirectional stretching sheet. The modeled partial differential equations are reduced in coupled ordinary differential equations by suitable similarity variables. The resulting equations are solved numerically by the shooting method using adaptive Runge Kutta algorithm in combination with Newton's method in the domain $[0,\infty)$. The numerical results for the velocity and temperature fields are graphically presented and effects of the relevant parameters are discussed in detail. Moreover, the skin-friction coefficient and local Nusselt number for different values of the power-law index and stretching ratio parameter are presented through tabulated data. The numerical results are also verified with the results obtained analytically by the homotopy analysis method (HAM). Additionally, the results are validated with previously published pertinent literature as a limiting case of the problem.*

1. **Introduction**

Recently, the surface driven flows have garnered considerable attention because of their practical importance in diverse engineering disciplines. Particularly, flows induced by the stretching surfaces are of significant importance in extrusion industry. The quality of the final products greatly depends upon the rate of heat transfer to ambient fluid. Various aspects of such problems have been the focal point of many theoretical researchers, since the pioneering work of Sakiadis [1]. Voluminous literature regarding the flow and heat transfer due to unidirectional stretching sheet is easily accessible. However, Wang [2] presented exact similar solutions for a three-dimensional flow due to stretching of sheet in two lateral directions. Later on, Ariel [3] addressed this problem by finding the approximate analytical solutions using the homotopy perturbation method. Liu and Anderson [4] also explored numerically the heat transfer characteristics of fluid, when the sheet is stretched in two lateral directions with variable thermal boundary conditions. Lakshmisha *et al.* [5] obtained numerical solutions of unsteady three-dimensional boundary layer with constant temperature and heat flux thermal boundary conditions. Ahmad *et al.* [6] provided the approximate analytical solutions of the problem for heat transfer in hydromagnetic flow induced by bidirectional stretching sheet in porous medium.

All of the abovementioned studies lie within the domain of a Newtonian fluid. There is a vast utility of non-Newtonian fluids in industrial sector such as pharmaceutical, polymer, personal care products and so forth. Schowalter [7] seems first who got the similar solutions for the boundary layer flow of power-law pseudoplastic fluid. Patil and Timol [8] found similarity solutions of three dimensional unsteady flow of power-law fluid past a flat plate. Substantial research has been carried out in the last decade on flow and heat transfer of non-Newtonian fluids over unidirectional stretching sheet, but scarce for bidirectional stretching sheet. Gorla *et al*. [9] addressed the three dimensional flow of a power-law fluid over a bidirectional stretching surface. Recently, Nadeem *et al*. [10] analyzed the MHD three-dimensional Casson fluid flow past a porous linearly stretching sheet. Shehzad *et al*. [11,12] also studied the three-dimensional flow of Maxwell and Oldroyd-B fluids over bi-dimensional stretching surface.

---

[1]Corresponding author: email: asifmunir1000@yahoo.com

To the best of our knowledge no attempt has been made so far to study the flow and heat transfer characteristics of Sisko fluid over a bidirectional stretching sheet. This paper aimed to analyze the unexplored flow and heat transfer characteristics in a boundary layer flow of Sisko fluid over a bidirectional stretching surface.

## 2. Physical model and mathematical formulation

### 2.1 Rheological model

In current study we consider time independent non-Newtonian fluid that follows Sisko rheological model. The Cauchy stress tensor **T** for such fluid is given by

$$\mathbf{T} = -p\mathbf{I} + \mathbf{S}, \tag{1}$$

where **S** is the extra stress tensor given by

$$\mathbf{S} = \left[ a + b \left| \sqrt{\frac{1}{2} tr(\mathbf{A}_1^2)} \right|^{n-1} \right] \mathbf{A}_1, \tag{2}$$

in which $n$, $a$ and $b$ are the physical constants different for different fluid, $\mathbf{A}_1 = (grad\mathbf{V}) + (gard\mathbf{V})^T$ the first Rivlin-Erickson tensor with **V** the velocity vector.

### 2.2 Boundary layer equations

The continuity and momentum equations for the steady flow of an incompressible Sisko fluid with the velocity field $\mathbf{V} = [u(x,y,z), v(x,y,z), w(x,y,z)]$ are stated as follows:

$$\frac{\partial u}{\partial x} + \frac{\partial v}{\partial y} + \frac{\partial w}{\partial z} = 0, \tag{3}$$

$$\rho\left(u\frac{\partial u}{\partial x} + v\frac{\partial u}{\partial y} + w\frac{\partial u}{\partial z}\right) = -\frac{\partial p}{\partial x} + a\left(\frac{\partial^2 u}{\partial x^2} + \frac{\partial^2 u}{\partial y^2} + \frac{\partial^2 u}{\partial z^2}\right) + 2b\frac{\partial}{\partial x}\left[\frac{\partial u}{\partial x}\left|\sqrt{\frac{1}{2}tr\mathbf{A}_1^2}\right|^{n-1}\right]$$
$$+ b\frac{\partial}{\partial y}\left[\left(\frac{\partial u}{\partial y} + \frac{\partial v}{\partial x}\right)\left|\sqrt{\frac{1}{2}tr\mathbf{A}_1^2}\right|^{n-1}\right] + b\frac{\partial}{\partial y}\left[\left(\frac{\partial u}{\partial z} + \frac{\partial w}{\partial x}\right)\left|\sqrt{\frac{1}{2}tr\mathbf{A}_1^2}\right|^{n-1}\right], \tag{4}$$

$$\rho\left(u\frac{\partial v}{\partial x} + v\frac{\partial v}{\partial y} + w\frac{\partial v}{\partial z}\right) = -\frac{\partial p}{\partial y} + a\left(\frac{\partial^2 v}{\partial x^2} + \frac{\partial^2 v}{\partial y^2} + \frac{\partial^2 v}{\partial z^2}\right) + b\frac{\partial}{\partial x}\left[\left(\frac{\partial u}{\partial y} + \frac{\partial v}{\partial x}\right)\left|\sqrt{\frac{1}{2}tr\mathbf{A}_1^2}\right|^{n-1}\right]$$
$$+ 2b\frac{\partial}{\partial y}\left[\left(\frac{\partial v}{\partial y}\right)\left|\sqrt{\frac{1}{2}tr\mathbf{A}_1^2}\right|^{n-1}\right] + b\frac{\partial}{\partial z}\left[\left(\frac{\partial v}{\partial z} + \frac{\partial w}{\partial y}\right)\left|\sqrt{\frac{1}{2}tr\mathbf{A}_1^2}\right|^{n-1}\right], \tag{5}$$

$$\rho\left(u\frac{\partial w}{\partial x}+v\frac{\partial w}{\partial y}+w\frac{\partial w}{\partial z}\right)=-\frac{\partial p}{\partial z}+a\left(\frac{\partial^2 w}{\partial x^2}+\frac{\partial^2 w}{\partial y^2}+\frac{\partial^2 w}{\partial z^2}\right)+b\frac{\partial}{\partial x}\left[\left(\frac{\partial u}{\partial z}+\frac{\partial w}{\partial x}\right)\left|\sqrt{\frac{1}{2}tr\mathbf{A}_1^2}\right|^{n-1}\right]$$

$$+b\frac{\partial}{\partial y}\left[\left(\frac{\partial v}{\partial z}+\frac{\partial w}{\partial y}\right)\left|\sqrt{\frac{1}{2}tr\mathbf{A}_1^2}\right|^{n-1}\right]+b\frac{\partial}{\partial z}\left[\frac{\partial w}{\partial z}\left|\sqrt{\frac{1}{2}tr\mathbf{A}_1^2}\right|^{n-1}\right], \quad (6)$$

where

$$\frac{1}{2}tr(\mathbf{A}_1^2)=2\left(\frac{\partial u}{\partial x}\right)^2+2\left(\frac{\partial v}{\partial y}\right)^2+2\left(\frac{\partial w}{\partial z}\right)^2+\left(\frac{\partial u}{\partial y}+\frac{\partial v}{\partial x}\right)^2+\left(\frac{\partial u}{\partial z}+\frac{\partial w}{\partial x}\right)^2+\left(\frac{\partial v}{\partial z}+\frac{\partial w}{\partial y}\right)^2. \quad (7)$$

Defining the dimensionless variables and parameters as

$$u^*=\frac{u}{U},\ v^*=\frac{v}{U},\ w^*=\frac{w}{U},\ x^*=\frac{x}{L},\ y^*=\frac{y}{L},\ z^*=\frac{z}{L},$$

$$p^*=\frac{p}{\rho U^2},\ \zeta_1=\frac{(a/\rho)}{LU}\ \text{and}\ \zeta_2=\frac{(b/\rho)}{LU}\left(\frac{U}{L}\right)^{n-1}, \quad (8)$$

where $L$ is the characteristic length and $U$ the characteristic velocity, The Eqs. (3) to (7) in terms of dimensionless variables and parameters can be casted as

$$\frac{\partial u^*}{\partial x^*}+\frac{\partial v^*}{\partial y^*}+\frac{\partial w^*}{\partial z^*}=0, \quad (9)$$

$$u^*\frac{\partial u^*}{\partial x^*}+v^*\frac{\partial u^*}{\partial y^*}+w^*\frac{\partial u^*}{\partial z^*}=-\frac{\partial p^*}{\partial x^*}+\zeta_1\left(\frac{\partial^2 u^*}{\partial x^{*2}}+\frac{\partial^2 u^*}{\partial y^{*2}}+\frac{\partial^2 u^*}{\partial z^{*2}}\right)+2\zeta_2\frac{\partial}{\partial x^*}\left[\frac{\partial u^*}{\partial x^*}\left|\sqrt{\frac{1}{2}tr\mathbf{A}_1^{*2}}\right|^{n-1}\right]$$

$$+\zeta_2\frac{\partial}{\partial y^*}\left[\left(\frac{\partial u^*}{\partial y^*}+\frac{\partial v^*}{\partial x^*}\right)\left|\sqrt{\frac{1}{2}tr\mathbf{A}_1^{*2}}\right|^{n-1}\right]+\zeta_2\frac{\partial}{\partial y^*}\left[\left(\frac{\partial u^*}{\partial z^*}+\frac{\partial w^*}{\partial x^*}\right)\left|\sqrt{\frac{1}{2}tr\mathbf{A}_1^{*2}}\right|^{n-1}\right], \quad (10)$$

$$u^*\frac{\partial v^*}{\partial x^*}+v^*\frac{\partial v^*}{\partial y^*}+w^*\frac{\partial v^*}{\partial z^*}=-\frac{\partial p^*}{\partial y^*}+\zeta_1\left(\frac{\partial^2 v^*}{\partial x^{*2}}+\frac{\partial^2 v^*}{\partial y^{*2}}+\frac{\partial^2 v^*}{\partial z^{*2}}\right)+\zeta_2\frac{\partial}{\partial x^*}\left[\left(\frac{\partial u^*}{\partial y^*}+\frac{\partial v^*}{\partial x^*}\right)\left|\sqrt{\frac{1}{2}tr\mathbf{A}_1^{*2}}\right|^{n-1}\right]$$

$$+2\zeta_2\frac{\partial}{\partial y^*}\left[\left(\frac{\partial v^*}{\partial y^*}\right)\left|\sqrt{\frac{1}{2}tr\mathbf{A}_1^{*2}}\right|^{n-1}\right]+\zeta_2\frac{\partial}{\partial z^*}\left[\left(\frac{\partial v^*}{\partial z^*}+\frac{\partial w^*}{\partial y^*}\right)\left|\sqrt{\frac{1}{2}tr\mathbf{A}_1^{*2}}\right|^{n-1}\right], \quad (11)$$

$$u^*\frac{\partial w^*}{\partial x^*}+v^*\frac{\partial w^*}{\partial y^*}+w^*\frac{\partial w^*}{\partial z^*}=-\frac{\partial p^*}{\partial z^*}+\zeta_1\left(\frac{\partial^2 w^*}{\partial x^{*2}}+\frac{\partial^2 w^*}{\partial y^{*2}}+\frac{\partial^2 w^*}{\partial z^{*2}}\right)+\zeta_2\frac{\partial}{\partial x^*}\left[\left(\frac{\partial u^*}{\partial z^*}+\frac{\partial w^*}{\partial x^*}\right)\left|\sqrt{\frac{1}{2}tr\mathbf{A}_1^{*2}}\right|^{n-1}\right]$$

$$+\zeta_2\frac{\partial}{\partial y^*}\left[\left(\frac{\partial v^*}{\partial z^*}+\frac{\partial w^*}{\partial y^*}\right)\left|\sqrt{\frac{1}{2}tr\mathbf{A}_1^{*2}}\right|^{n-1}\right]+\zeta_2\frac{\partial}{\partial z^*}\left[\frac{\partial w^*}{\partial z^*}\left|\sqrt{\frac{1}{2}tr\mathbf{A}_1^{*2}}\right|^{n-1}\right], \quad (12)$$

$$\frac{1}{2}tr(\mathbf{A}_1^{*2})=2\left(\frac{\partial u^*}{\partial x^*}\right)^2+2\left(\frac{\partial v^*}{\partial y^*}\right)^2+2\left(\frac{\partial w^*}{\partial z^*}\right)^2+\left(\frac{\partial u^*}{\partial y^*}+\frac{\partial v^*}{\partial x^*}\right)^2+$$

$$\left(\frac{\partial u^*}{\partial z^*}+\frac{\partial w^*}{\partial x^*}\right)^2+\left(\frac{\partial v^*}{\partial z^*}+\frac{\partial w^*}{\partial y^*}\right)^2. \quad (13)$$

Now applying the order of magnitude arguments to ignore the small terms, keeping in view that the dimensionless parameters $\zeta_1$ and $\zeta_2$ are of the order $\delta^2$ and $\delta^{n+1}$, with $\delta \ll 1$. Within the framework of these assumptions Eqs. (9) to (13) in dimensional form simplifies as:

$$\frac{\partial u}{\partial x} + \frac{\partial v}{\partial y} + \frac{\partial w}{\partial z} = 0, \tag{14}$$

$$\rho\left(u\frac{\partial u}{\partial x} + v\frac{\partial u}{\partial y} + w\frac{\partial u}{\partial z}\right) = -\frac{\partial p}{\partial x} + a\frac{\partial^2 u}{\partial z^2} + b\frac{\partial}{\partial z}\left(\left|\frac{\partial u}{\partial z}\right|^{n-1}\frac{\partial u}{\partial z}\right), \tag{15}$$

$$\rho\left(u\frac{\partial v}{\partial x} + v\frac{\partial v}{\partial y} + w\frac{\partial v}{\partial z}\right) = -\frac{\partial p}{\partial y} + a\frac{\partial^2 v}{\partial z^2} + b\frac{\partial}{\partial z}\left(\left|\frac{\partial u}{\partial z}\right|^{n-1}\frac{\partial v}{\partial z}\right). \tag{16}$$

*2.3 Governing equations and boundary conditions*

Considering the three-dimensional steady, laminar and incompressible flow of fluid obeying the Sisko model which occupy space $z > 0$. The fluid is set into motion by an elastic flat sheet in the plane $z = 0$ kept at a constant temperature, which is being continuously stretched with linear velocities *cx* and *dy* in the *x*- and *y*- directions, respectively. The ambient temperature far away from the sheet is taken as $T_\infty$. The continuity, momentum and energy equations governing the steady three-dimensional flow of Sisko fluid, approximated by boundary-layer theory, are

$$\frac{\partial u}{\partial x} + \frac{\partial v}{\partial y} + \frac{\partial w}{\partial z} = 0, \tag{17}$$

$$\rho\left(u\frac{\partial u}{\partial x} + v\frac{\partial u}{\partial y} + w\frac{\partial u}{\partial z}\right) = a\frac{\partial^2 u}{\partial z^2} - b\frac{\partial}{\partial z}\left(-\frac{\partial u}{\partial z}\right)^n, \tag{18}$$

$$\rho\left(u\frac{\partial v}{\partial x} + v\frac{\partial v}{\partial y} + w\frac{\partial v}{\partial z}\right) = \frac{\partial^2 v}{\partial z^2} + b\frac{\partial}{\partial z}\left(-\frac{\partial u}{\partial z}\right)^{n-1}\frac{\partial v}{\partial z}, \tag{19}$$

$$u\frac{\partial T}{\partial x} + v\frac{\partial T}{\partial y} + w\frac{\partial T}{\partial z} = \frac{\kappa}{\rho c_p}\frac{\partial^2 T}{\partial z^2}. \tag{20}$$

Here $(u, v, w)$ are velocity components, $T$ the temperature, $\rho$ the fluid density, $c_p$ the specific heat of fluid at constant pressure and $\kappa$ the thermal conductivity.

Equations (17) to (20) are subjected to the following boundary conditions

$$u = U_w(x) = cx,\ v = V_w(y) = dy,\ w = 0,\ T = T_w \text{ at } z = 0, \tag{21}$$

$$u \to 0, v \to 0, w \to 0,\ T \to T_\infty \text{ as } z \to \infty. \tag{22}$$

## 2.4 Transformed problem

The governing coupled partial differential equations (18) to (20) are transformed to coupled ordinary differential equations by introducing similarity variables

$$u = cxf'(\eta), \ v = dyg'(\eta), \ w = -c\left(\frac{c^{n-2}}{\rho/b}\right)^{1/(n+1)}\left(\frac{2n}{n+1}f + \frac{1-n}{1+n}\eta f' + g\right)x^{(n-1)/(n+1)};$$

$$\theta(\eta) = \frac{T(x,y,z) - T_\infty}{T_w - T_\infty}, \eta = z\left(\frac{c^{2-n}}{b/\rho}\right)^{1/n+1} x^{(1-n)/(1+n)}. \tag{23}$$

The momentum and heat transfer equations with the associated boundary conditions reduces to

$$Af''' + n[-f'']^{n-1} f''' + \frac{2}{n+1} ff'' - (f')^2 + gf'' = 0, \tag{24}$$

$$Ag''' + (-f'')^{n-1} g''' - (n-1) g''f'''(-f'')^{n-2} + \frac{2n}{n+1} fg'' - (g')^2 + gg'' = 0, \tag{25}$$

$$\theta'' + \Pr\left(\frac{2n}{n+1}\right) f\theta' + \Pr g\theta' = 0, \tag{26}$$

$$f(0) = 0, \ g(0) = 0, \ f'(0) = 1, \ g'(0) = c/d = \alpha, \ \theta(0) = 1, \tag{27}$$

$$f'(\eta) \to 0, \ g'(\eta) \to 0 \text{ and } \theta(\eta) \to 0 \text{ as } \eta \to \infty, \tag{28}$$

where the primes stand for differentiation with respect $\eta$ and $\alpha \in [0,1]$ is the stretching ratio parameter. Further, $\text{Re}_a$, $\text{Re}_b$ are the local Reynolds number, $A$ the material parameter of Sisko fluid and Pr the generalized Prandtl number, which are defined as

$$\text{Re}_a = \frac{\rho xU}{a}, \ \text{Re}_b = \frac{\rho x^n U^{2-n}}{b}, \ A = \frac{\text{Re}_b^{\frac{2}{n+1}}}{\text{Re}_a} \text{ and } \Pr = \frac{xUR_b^{\frac{-2}{n+1}}}{\kappa/\rho c_p}. \tag{29}$$

Note that in the limit $\alpha \to 0$ the unidirectional case is obtained and motion of fluid is merely in $xz$ plane, i.e. $g$ and $g'$ in Eq. (23) both are zero. When, $\alpha = 1$, the stretching rate is same in the $x$- and $y$- directions and the flow is axisymmetric.

## 2.5 Important physical parameters

The physical quantities of main interest are the skin-friction coefficient and the local Nusselt number.

*2.5.1 The skin-friction coefficients*

The skin coefficient is an important boundary layer characteristic, which is the dimensionless shear stress at the wall $(z=0)$. Thus the dimensionless skin friction coefficients along the $x$- and $y$-directions, respectively, are given by

$$C_{fx} = \frac{\tau_{xz}}{\frac{1}{2}\rho U_w^2} \text{ and } C_{fy} = \frac{\tau_{yz}}{\frac{1}{2}\rho U_w^2}, \qquad (30)$$

where $\tau_{xz}$ and $\tau_{yz}$ are shear stresses along the $x$- and $y$- directions, respectively. These quantities in dimensionless form can be expressed as

$$\frac{1}{2}\text{Re}_b^{1/(n+1)} C_{fx} = Af''(0) + [f''(0)]^n, \qquad (31)$$

$$\frac{1}{2}\text{Re}_b^{1/(n+1)} C_{fy} = \frac{U_w}{V_w}\left[Ag''f(0) + [f''(0)]^{n-1} g''(0)\right]. \qquad (32)$$

*2.5.2 The local Nusselt number*

The local Nusselt number denoted by $Nu_x$, giving the rate of heat transfer at the wall, is defined by

$$Nu_x = \frac{xq_w}{\kappa(T_w - T_\infty)}\bigg|_{y=0}, \qquad (33)$$

where the wall heat flux defined by $q_w = -\kappa\left(\frac{\partial T}{\partial y}\right)\bigg|_{y=0}$ resulting in

$$\text{Re}_b^{-1/n+1} Nu_x = -\theta'(0). \qquad (34)$$

## 3. Solution methodologies

*3.1 Numerical solution method*

We are not able to find the exact analytical solution of the non-linear two point boundary value problem (24) - (28). Consequently, these equations are solved numerically by the shooting technique. The equations are written as a system of eight first order ordinary differential equations. The corresponding initial value problem is solved by adaptive Runge-Kutta method. The Newton's iterative algorithm is used to assure quadratic convergence of iteration required to satisfy the boundary condition at infinity.

*3.2 Analytical solution method*

Analytical results are sought using the homotopy analysis method (HAM) for certain values of parameters with a view to check the veracity of our numerical results. To apply the HAM, we selected the initial guesses agreeing the boundary data as

$$f_0 = 1 - e^{-\eta}, \; g_0 = \lambda(1 - e^{-\eta}) \text{ and } \theta_0 = e^{-\eta}, \qquad (35)$$

and auxiliary linear operators
$$L_f = f''' - f', L_g = g''' - g' \text{ and } L_\theta = \theta'' - \theta, \tag{36}$$
for the velocity and temperature fields, respectively.

The proper values of convergence control parameters $\hbar_f, \hbar_g$ and $\hbar_\theta$ assure the convergence of series solution. The optimal values of these parameters are chosen by minimizing the discrete squared residual error [13].

## 4. Results and discussion

This article predominantly focuses the flow and heat transfer characteristics of Sisko fluid past a uniformly heated and bidirectional stretching surface. To grasp both the phenomena, Eqs. (24) to (26) with the boundary conditions (27) and (28) are solved numerically for non-integer values of the power-law index $n$ and results are presented graphically. The effects of various parameters like the power-law index $n$, material parameter $A$ and stretching ratio parameter $\alpha$ are investigated for non-dimensional velocity components $f'(\eta)$, $g'(\eta)$ and temperature $\theta(\eta)$. Variation of the local skin friction coefficients and local Nusselt number are also observed for different values of $n$ and $\alpha$.

Figures 1 $(a,b,c,d)$ present the effect of the stretching ratio parameter $\alpha$ on the velocity components $f'(\eta)$ and $g'(\eta)$ for Sisko fluid with shear thinning ($n=0.5$) and shear thickening ($n=1.5$) properties. It is evident from figure 1(a) that $f'(\eta)$ decreases with an increase in $\alpha$, however, opposite behavior is observed for the $g'(\eta)$ component of the velocity (figure 1b). A comparison of figures 1 $(a,b)$ reveals that the stretching ratio parameter affects $g'(\eta)$ component more pronouncedly. Figures 1 $(c,d)$ show the same qualitative trends but there is noticeable decrease in boundary layer thickness for shear thickening fluid as compared to shear thinning fluid.

Figures 2$(a,b,c,d)$ depict how the material parameter $A$ affects the velocity profiles $f'(\eta)$ and $g'(\eta)$ for Sisko fluid with shear thinning and shear thickening properties. Figures 2$(a,b)$ show that the velocities profiles increases monotonically with each increment of $A$. The rest of the figures show the same qualitative behavior. Again, figures 2$(c,d)$ reveal that there is marked decrease in the boundary layer thickness for shear thickening fluids.

Figures 3$(a,b)$ depict the influence of stretching ratio parameter $\alpha$ on dimensionless temperature $\theta(\eta)$. From these figures a decrease in $\theta(\eta)$ is noticed for each increment in value of $\alpha$. The effect is more noticable for shear thinning fluid (figure 3a). Further, it is observed that the thermal boundary layer thickness decreases for shear thickening ($n=1.5$) fluid.

The following figures of temperature profiles are plotted and discussed for selected values of stretching ratio parameter $\alpha$, the Prandtl number Pr and the material parameter $A$ of the Sisko fluid. Figures 4$(a,b,c,d)$ elucidate the effect of the Prandtl number Pr on temperature profile $\theta(\eta)$ for fixed values of the power-law index $n$ and the stretching ratio parameter $\alpha$. The figures reveal that the temperature profile decreases when Pr is increased. Further, figures 4$(b,d)$ exhibit that there is a decrease in thermal boundary layer thickness at higher value of the stretching ratio parameter $\alpha$.

To exhibit the effects of the material parameter $A$ of Sisko fluid on non-dimensional temperature profile, we have plotted figures 5(a,b). These plots clearly show that the temperature profile decreases monotonically with increasing values of $A$ for the shear thinning and shear thickening fluids. Again, it is noticed that the thermal boundary layer thickness is markedly reduced for shear-thickening ($n=1.5$) fluid. These figures also provide a comparison between the profiles of the power-law fluid ($A=0$) with those of the Sisko fluid ($A \neq 0$). From these figures, it is clear that the temperature profiles for the power law fluid is higher when compared with those of Sisko fluid.

The veracity of the numerical results are confirmed by comparison with the analytical results obtained by HAM. Additionally, the results are compared with previous published relevant literature (table 1) as a special case of the problem and excellent agreement is observed in both the cases.

The magnitude of the numerical values of the skin friction coefficients in the $x$ and $y$ directions are given in table 2. It is clear from table that the skin friction drag is higher to some extent for the power-law index less than one. Moreover, it increases with an increase in value of the stretching ratio parameter $\alpha$. Further, it is apparent that the increase is more rapid for the power-law index $n>1$. Table 3 presents the variation in wall temperature gradient $-\theta'(\eta)$ with the stretching ratio parameter $\alpha$. It is observed that the value of $-\theta'(\eta)$ increases with each increment of $\alpha$. This table depicts that the wall temperature gradient is higher for fluid with shear thickening properties and hence results in a higher heat transfer coefficient.

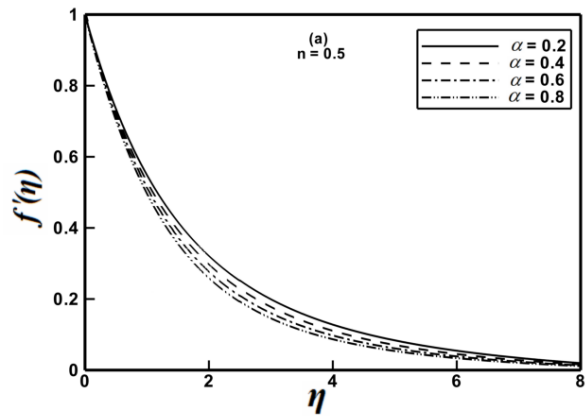
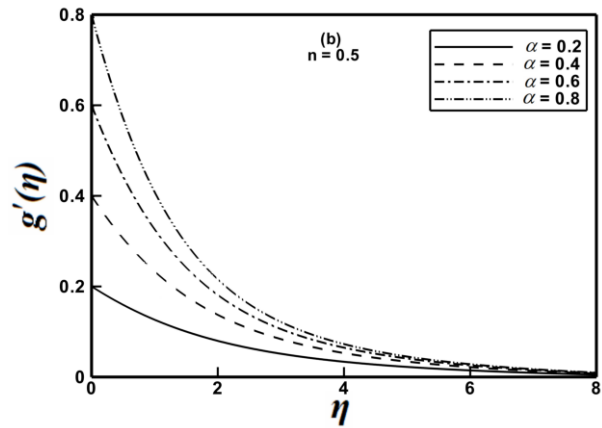

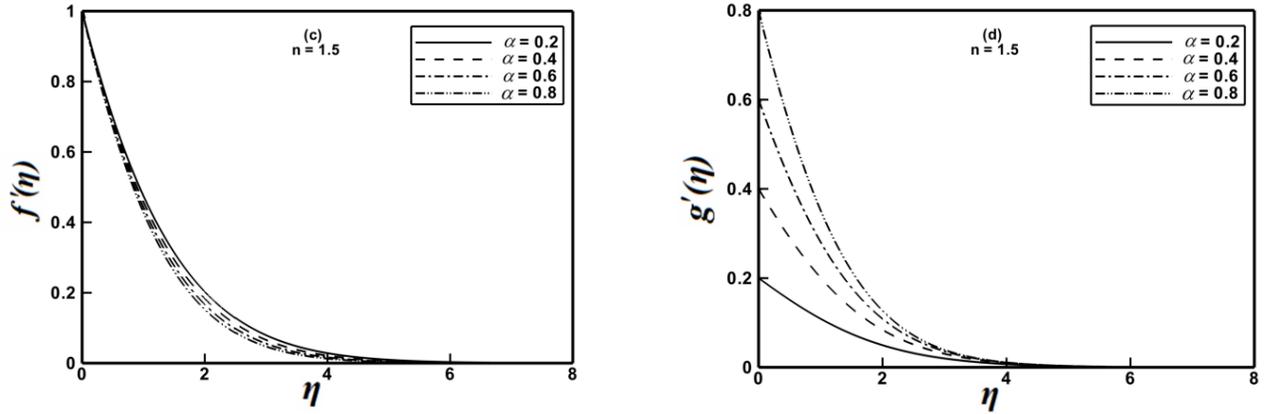

Figure 1: Profiles of the velocities $f'(\eta)$ and $g'(\eta)$ for different values of the stretching ratio parameter $\alpha$ when $A = 1.5$ is fixed.

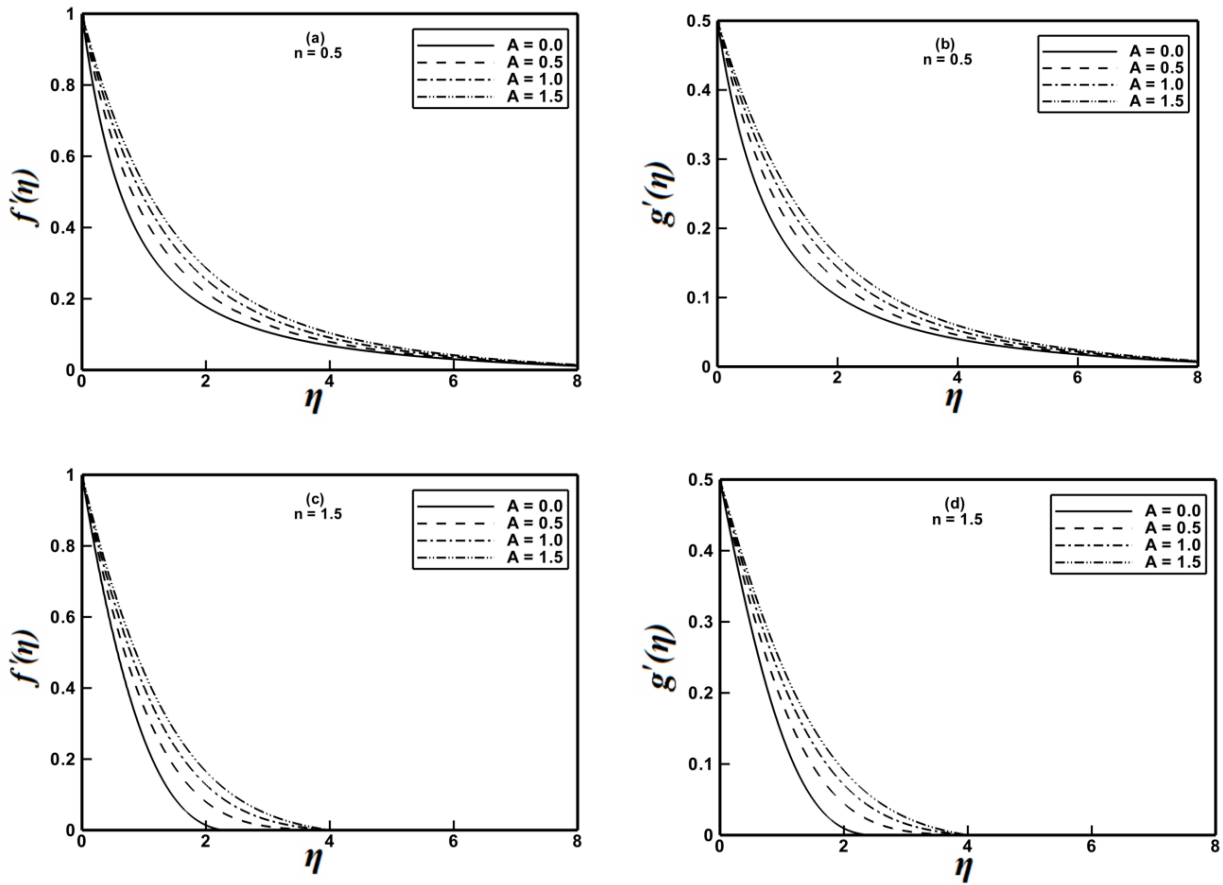

Figure 2: Profiles of the velocities $f'(\eta)$ and $g'(\eta)$ for different values of the material parameter $A$ when $\alpha = 0.5$ is fixed.

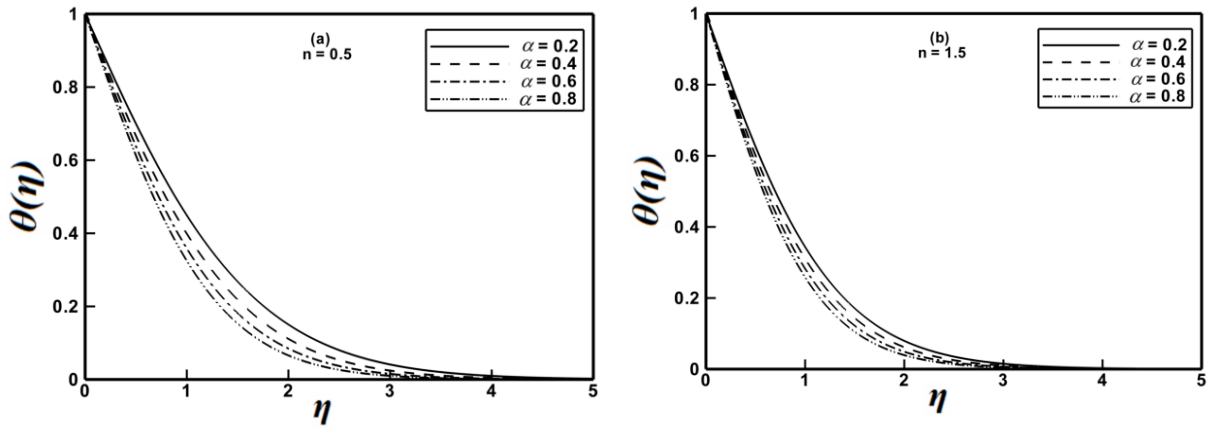

Figure 3: Profiles of the temperature $\theta(\eta)$ for different values of the stretching ratio parameter $\alpha$ when $A=1.5$ and $\Pr=1.0$ are fixed.

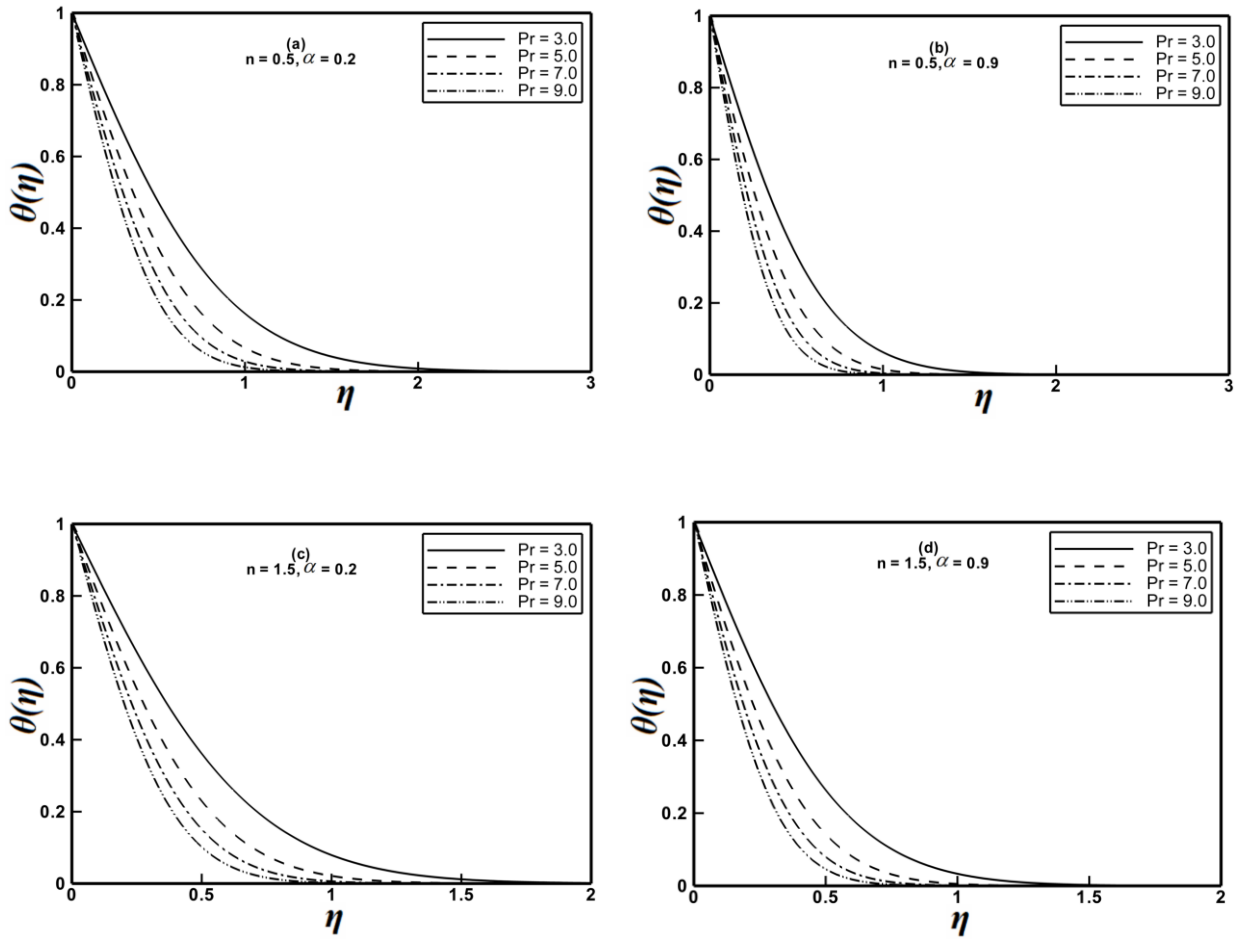

Figure 4: Profiles of the temperature $\theta(\eta)$ for different values of the Prandtl number Pr when $A=1.5$ is fixed.

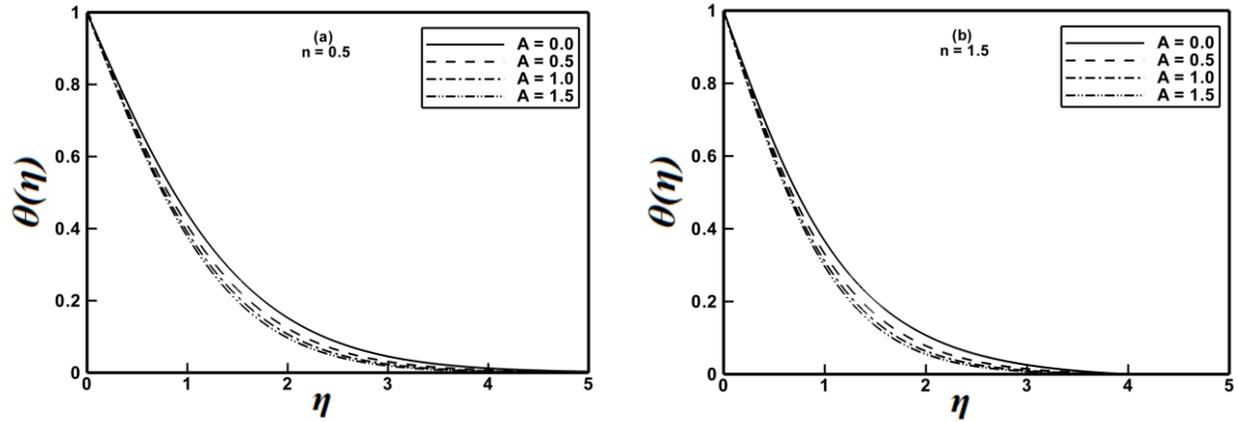

figure 5: Profiles of the temperature $\theta(\eta)$ for different values of the material parameter $A$ when $\alpha = 0.5$ is fixed.

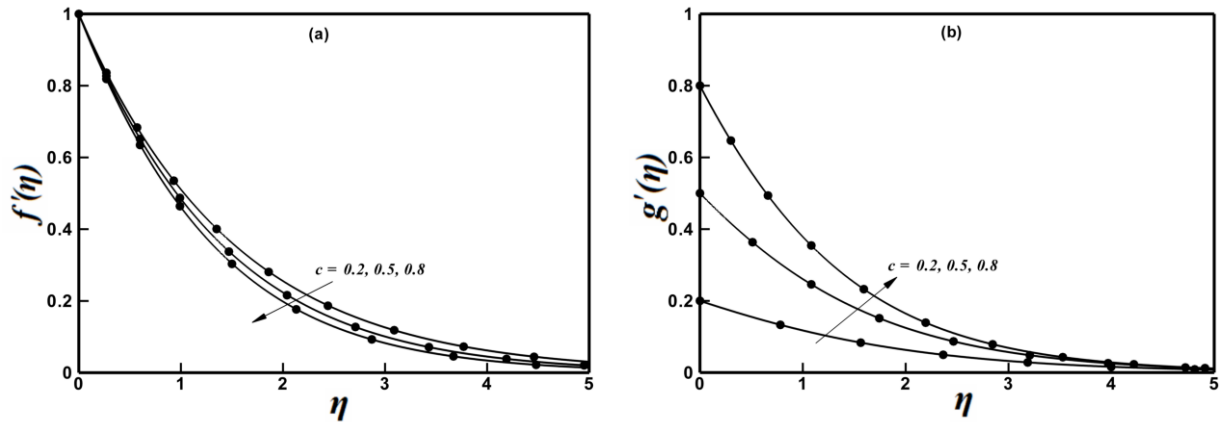

Figure 6: A comparison between the numerical and HAM results (solid lines numerical and solid circles HAM results) when $n=1.0$ and $A=1.5$ are fixed.

Table 1: A comparison of the local skin friction coefficient and local Nusselt number when $n=1$ and $A=0$ are fixed.

|  |  | $f''(0)$ | $g''(0)$ | $\theta'(0)$ |
|---|---|---|---|---|
| Wang [2] |  | -1.00000 | 0 | - |
| Lui et al.[4] | $\alpha = 0.0$ | -1.00000 | 0 | - |
| Present study |  | -1.00000 | 0 | - |
| Wang [2] |  | -1.048813 | -0.194564 | - |
| Lui et al.[4] | $\alpha = 0.25$ | -1.048813 | -0.194565 | -0.665933 |
| Present study |  | -1.048818 | -0.194567 | -0.665939 |
| Wang [2] |  | -1.093097 | -0.465205 | - |
| Lui et al.[4] | $\alpha = 0.50$ | -1.093096 | -0.465206 | -0.735334 |
| Present study | | -1.093098 | -0.465207 | -0.735336 |
| Wang [2] | | -1.134485 | -0.794622 | - |
| Lui et al.[4] | $\alpha = 0.75$ | -1.134486 | -0.794619 | -0.796472 |
| Present study | | -1.134487 | -0.794619 | -0.796472 |
| Wang [2] | | -1.173720 | -1.173720 | - |
| Lui et al.[4] | $\alpha = 1.0$ | -1.173721 | -1.173721 | - |
| Present study | | -1.173721 | -1.173721 | - |

Table 2: A tabulation of the local skin friction coefficients along the *x*- and *y*- directions.

| | $n = 0.5$ | | $n = 1.5$ | |
|---|---|---|---|---|
| $\alpha$ | $\frac{1}{2}\text{Re}_b^{1/(n+1)} C_{fx}$ | $\frac{1}{2}\text{Re}_b^{1/(n+1)} C_{fy}\left(\frac{V_w}{U_w}\right)$ | $\frac{1}{2}\text{Re}_b^{1/(n+1)} C_{fx}$ | $\frac{1}{2}\text{Re}_b^{1/(n+1)} C_{fy}\left(\frac{V_w}{U_w}\right)$ |
| 0.2 | -1.74610 | -0.24536 | -1.60218 | -0.23549 |
| 0.4 | -1.79819 | -0.57842 | -1.66050 | -0.55234 |
| 0.6 | -1.84739 | -0.97856 | -1.71569 | -0.93386 |
| 0.8 | -1.89434 | -1.43463 | -1.76846 | -1.37047 |

Table 3: A tabulation of the local Nusselt number.

| | $\text{Re}_b^{-1/(n+1)} Nu_x$ | |
|---|---|---|
| $\alpha$ | $n = 0.5$ | $n = 1.5$ |
| 0.2 | -0.62074 | -0.78919 |
| 0.4 | -0.69468 | -0.84864 |
| 0.6 | -0.75957 | -0.90287 |
| 0.8 | -0.81827 | -0.95324 |

## 5. Conclusions

The steady three-dimensional flow and heat transfer characteristics within boundary layer of Sisko fluid past an isothermal bidirectional stretching surface has been studied numerically. The effects of the stretching ratio parameters, the material parameter and the Prandtl number on the velocity and temperature profiles were studied. Our computations have indicated that:

- A qualitatively opposite trend was observed in the velocity components $f'(\eta)$ and $g'(\eta)$ for increasing value of the stretching ratio parameter.
- A substantial reduction in the momentum and thermal boundary layer thickness was noticed

for shear thickening fluid.

- Thinning of the thermal boundary layer was seen for increasing Prandtl number and stretching ratio parameter and hence resulting in better heat transfer.
- The skin friction drag and wall temperature gradient was increased with the stretching ratio parameter.

**Acknowledgement**

This work has the financial support of the Higher Education Commission (HEC) of Pakistan.